\documentclass[prd,a4paper,nofootinbib]{revtex4}
\usepackage{epsfig,graphicx}
\newcommand{\be}{\begin{equation}}
\newcommand{\ee}{\end{equation}}
\newcommand{\bea}{\begin{eqnarray}}
\newcommand{\eea}{\end{eqnarray}}
\newcommand{\bean}{\begin{eqnarray*}}
\newcommand{\eean}{\end{eqnarray*}}

\newcommand{\gapproxeq}{\lower
.7ex\hbox{$\;\stackrel{\textstyle >}{\sim}\;$}}
\newcommand{\lapproxeq}{\lower
.7ex\hbox{$\;\stackrel{\textstyle <}{\sim}\;$}}

\begin{document}

\bibliographystyle{unsrt}

\title{\bf Further insights into the ``$\rho\pi$ puzzle" }

\author{Qiang Zhao$^{1,2,3}$ and Gang Li$^1$}
\affiliation{1) Institute of High Energy Physics, Chinese Academy of
Sciences, Beijing, 100049, P.R. China}

\affiliation{2) Department of Physics, University of Surrey,
Guildford, GU2 7XH, United Kingdom}

\affiliation{3) Theoretical Physics Center for Science Facilities,
Chinese Academy of Sciences, Beijing 100049, P.R. China}

\author{Chao-Hsi Chang}
\affiliation{Institute of Theoretical Physics, Chinese Academy of
Sciences, Beijing, 100080, P.R. China}

\date{\today}

\begin{abstract}

Based on a systematic investigation of $J/\psi(\psi^\prime)\to VP$,
where $V$ and $P$ stand for light vector and pseudoscalar mesons, we
identify the role played by the electromagnetic (EM) transitions and
intermediate meson loop transitions, which are essential ingredients
for understanding the $J/\psi$ and $\psi^\prime$ couplings to $VP$.
We show that on the one hand, the EM transitions have relatively
larger interferences in $\psi^\prime\to \rho\pi$ and
$K^*\bar{K}+c.c.$ as explicitly shown by vector meson dominance
(VMD). On the other hand, the strong decay of $\psi^\prime$ receives
relatively larger destructive interferences from the intermediate
meson loop transitions. By identifying these mechanisms in an
overall study of $J/\psi(\psi^\prime)\to VP$, we provide a coherent
understanding of the so-called ``$\rho\pi$ puzzle".

\end{abstract}

\maketitle





\vspace{1cm}

\section{Introduction}

The decay channels $J/\psi (\psi^\prime) \to VP$, which are
suppressed in QCD due to the violation of hadronic-helicity
conservation~\cite{Brodsky:1981kj}, have attracted much attention in
the past few decades. According to this selection rule, one expects
the ratio $BR(\psi^\prime\to \rho\pi)/BR(J/\psi\to \rho\pi)\simeq
(M_{J/\psi}/M_\psi^\prime)^6 \sim 0.35$~\cite{Brodsky:1981kj}, which
turns out to be much larger than the experimental
data~\cite{pdg2006}, $BR(\psi^\prime\to \rho\pi)/BR(J/\psi\to
\rho\pi)\simeq (0.2\pm 0.1)\%$. This significant discrepancy is
known as the so-called ``$\rho\pi$ puzzle". An alternative
expression of the ``$\rho\pi$ puzzle" is via the ratios between
$J/\psi$ and $\psi^\prime$ annihilating into three gluons and a
single direct photon: \bea R &\equiv & \frac{BR(\psi^\prime\to
hadrons)}{BR(J/\psi\to hadrons)} \simeq \frac{BR(\psi^\prime\to e^+
e^-)}{BR(J/\psi\to e^+ e^-)}\simeq 12\% , \eea which is empirically
called ``12\% rule". The puzzle arises from the violation of the
above empirical rule in exclusive channels such as $\rho\pi$ and
$K^*\bar{K}+c.c.$, where the branching ratio fractions are found to
be orders-of-magnitude smaller than the approximate ``12\%".

Since the first observation of such a large deviation by Mark-II
Collaboration in 1983~\cite{Franklin:1983ve}, many theoretical
explanations have been proposed to decipher this
puzzle~\cite{Hou:1982kh,Karl:1984en,Pinsky:1984da,Brodsky:1987bb,Chaichian:1988kn,Pinsky:1989ue,Brodsky:1997fj,Li:1996yn,Chen:1998ma,gerald,Feldmann:2000hs,Suzuki:2001fs,Rosner:2001nm,Wang:2003zxa,Liu:2006dq}.
They can be classified into three categories: i)
$J/\psi$-enhancement hypothesis, which attributes the small
$R$-value to the enhanced branching fraction of $J/\psi$ decays; ii)
$\psi^\prime$-suppression hypothesis, which attributes the small
$R$-value to the small branching ratio of  $\psi^\prime$ decays;
iii) and other hypotheses which do not simply belong to the above
two categories. Unfortunately, so far none of those solutions has
been indisputably agreed~\cite{yuan,m-y-w}.

In this proceeding, we report our recent efforts on understanding
this issue. We shall identify i) the role play by EM transition in
$J/\psi (\psi^\prime) \to VP$ in a VMD model, which has relatively
large interferences in $\rho\pi$ and $K^*\bar{K}+c.c.$ channel; and
ii) mechanisms which suppress the strong decay amplitudes for
$\psi^\prime\to VP$. We emphasize that our analysis is based on so
far the-state-of-art experimental measurements of
$J/\psi(\psi^\prime)\to VP$~\cite{pdg2006}. The systematic exposed
in this study can  provide some insights into the charmonium
hadronic decays on a much more general ground.

\section{Step 1: EM transitions in the VMD model}

The importance of EM transitions in $J/\psi(\psi^\prime)\to VP$ can
be recognized by explicit experimental observations. For instance,
the branching ratios for isospin-violating decays,
$J/\psi(\psi^\prime)\to \rho\eta$, $\rho\eta^\prime$, etc are
compatible with those isospin-conserving channels such as
$\omega\eta$, $\omega\eta^\prime$, and $\phi\eta$
etc~\cite{pdg2006}. This is an evidence showing that the strong
decay amplitudes become suppressed and have the same order of
magnitude as the isospin-violating amplitudes, i.e. EM and strong
isospin-violating transition.

\begin{figure}
\begin{center}
\includegraphics[scale=0.6]{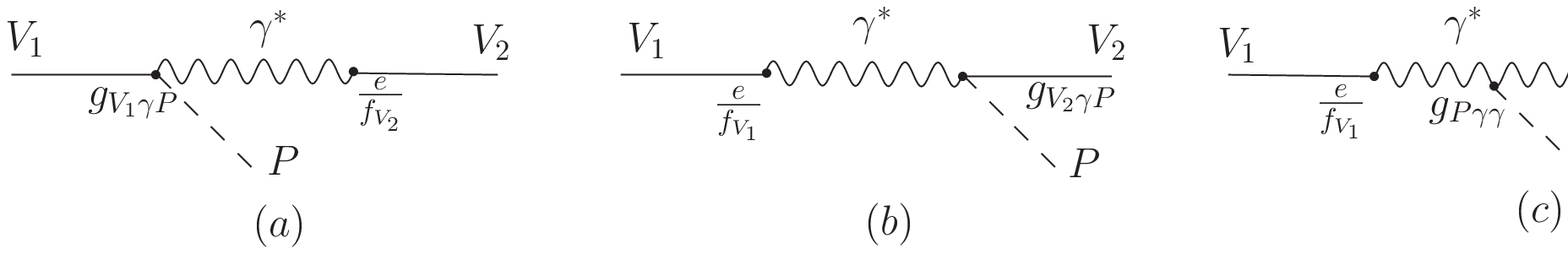}
\caption{Schematic diagrams for $J/\psi (\psi^\prime)\to \gamma^*\to
VP$. } \protect\label{fig-1}
\end{center}
\end{figure}

The role of the EM can be separately investigated due to the
available experimental data for vector meson radiative decays, i.e.
$\omega$, $\rho$, $\phi$, $K^*$, $J/\psi$ and
$\psi^\prime$~\cite{pdg2006}. Moreover, precise measurements of
vector meson decays into lepton pairs such as $e^+ e^-$ are also
available. This allows a well-constraint on the coupling constants
required in the VMD model, and only leaves an overall form factor
which takes care of the off-shell couplings, to be determined by
experimental data. Three independent EM transitions for $V_1\to V_2
P$ are illustrated by Fig.~\ref{fig-1}.


Typical effective Lagrangian for the $V\gamma P$ coupling are:
\be\label{lagrangian-1} {\cal L}_{V\gamma P}=\frac{g_{V\gamma
P}}{M_V}\epsilon_{\mu\nu\alpha\beta}\partial^\mu
V^\nu\partial^\alpha A^\beta P \ee where $V^\nu(=\rho, \ \omega, \
\phi, \ J/\psi, \ \psi^\prime\dots)$ and $A^\beta$ are the vector
meson and EM field, respectively; $M_V$ is the vector meson mass;
$\epsilon_{\mu\nu\alpha\beta}$ is the anti-symmetric Levi-Civita
tensor.

The $V\gamma^*$ coupling is described in VMD model, \be {\cal
L}_{V\gamma}=\sum_V \frac{e M_V^2}{f_V} V_\mu A^\mu \ , \ee where
$eM_V^2/f_V$ is a direct photon-vector-meson coupling in Feynman
diagram language, and the isospin 1 and 0 component of the EM field
are both included.

The invariant transition amplitude for $V_1\to\gamma^*\to  V_2 P$
can thus be expressed as: \bea \label{4}
{\cal M}_{EM} & \equiv & {\cal M}_A + {\cal M}_B + {\cal M}_C \nonumber\\
&=& \left( \frac{e}{f_{V2}}\frac{g_{V1\gamma P}}{M_{V1}}{\cal F}_a +
\frac{e}{f_{V1}}\frac{g_{V2\gamma P}}{M_{V2}}{\cal F}_b +
\frac{e^2}{f_{V1} f_{V2}} \frac{g_{P\gamma\gamma}}{M_P}{\cal F}_c
\right)\epsilon_{\mu\nu\alpha\beta}\partial^\mu
V_1^\nu\partial^\alpha V_2^\beta P \eea where $g_{P\gamma\gamma}$ is
the coupling for the neutral pseudoscalar meson decay to two
photons; ${\cal F}_a$, ${\cal F}_b$ and ${\cal F}_c$ denote the form
factor corrections to the transition of figure~\ref{fig-1}. A
monopole (MP) form factor is adopted here,
 \be \label{mp}{\cal
F}(q^2) = \frac {1} {1-q^2/\Lambda^2},
 \ee
 with $\Lambda = 0.542\pm
0.008$ GeV and $\Lambda = 0.577\pm 0.011$ GeV determined by the
isospin violated channels  $J/\psi (\psi^\prime) \to \rho\eta$,
$\rho\eta^\prime$, $\omega\pi^0$, and $\phi\pi^0$ with a
constructive (MP-C) or destructive phase (MP-D) between
Fig.~\ref{fig-1} (a) and (b), respectively. The form factors are
introduced because we think that the non-perturbative QCD effects
may play an important role in the transition at $J/\psi$ energy
scale.

The form factor ${\cal F}_c$ appearing in Eqs.~(\ref{4}) can be
determined in $\gamma^*\gamma^*$ scatterings. A commonly adopted
form factor is \be {\cal
F}_c(q_1^2,q_2^2)=\frac{1}{(1-q_1^2/\Lambda^2)(1-q_2^2/\Lambda^2)} \
, \ee where $q_1^2=M_{V1}^2$ and $q_2^2=M_{V2}^2$ are the squared
four-momenta carried by the time-like photons. We assume that the
$\Lambda$ is the same as in Eq.~(\ref{mp}), thus, ${\cal F}_c={\cal
F}_a {\cal F}_b$.

It should be noted that in Fig.~\ref{fig-1} the direct application
of $V\gamma P$ couplings extracted from experimental data will avoid
uncertainties arising from a $\gamma \to V^\prime\to VP$ treatment.
Unknown energy-dependence of those couplings can then be absorbed
into an overall form factor ${\cal F}(q^2)$ for which the cut-off
energy is determined by fitting those isospin-violating decay
branching ratios, i.e. $J/\psi(\psi^\prime) \to
\rho\eta,~\rho\eta^\prime,~\omega\pi$ and $\phi\pi$.

In fact, one can learn more from the isospin-violating channels. If
the EM transition is the dominant transition mechanism, one can
expect that the $12\%$ will be reasonably respected given that the
$J/\psi$ and $\psi^\prime$ wavefunctions are normal $c\bar{c}$ of
$(1S)$ and $(2S)$, respectively, and no significant interferences
from other processes. As shown in Table~\ref{tab-1}, one indeed sees
that the $12\%$ rule is satisfied though the experimental values
still have large uncertainties. There might be contributions from
strong isospin-violating transitions. However, the present
experimental results suggest that their interferences with the EM
transitions in the isospin-violating channel are relatively small.

 \begin{table}[ht]
 \begin{tabular}{ccc}
 \hline
 Decay channels     & $R^{VP}(\%)$   & Exp.data (\%)  \\[1ex]\hline
 $\rho\eta$         &  {8.97}                                      & $11.5\pm 5.0 $\\[1ex]
 $\rho\eta^\prime$  & {9.44}                                      & $23.5\pm 17.8$ \\[1ex]
 $\omega\pi$        &   {9.01}                                      & $5.0\pm 1.8$\\[1ex]
 $\phi\pi$          &   {7.41}                                      & $< 62.5 $ \\[1ex]\hline
\end{tabular}
\caption{  Branching ratio fractions of $\psi^\prime\to\gamma^*\to
VP$ over $J/\psi\to\gamma^*\to VP$ for those isospin-violating
channels. Here, we only show results with MP-C form factor. The last
column is extracted from the experimental date~\cite{pdg2006}.}
\label{tab-1}
\end{table}

\section{Step 2: Parameterize the strong decay transitions}

\begin{figure}
\begin{center}
\includegraphics[scale=0.8]{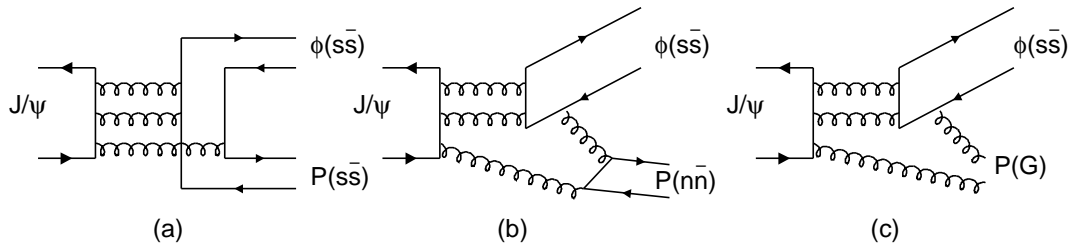} \caption{Schematic
diagrams for $J/\psi \to \phi P$ via strong interaction, where the
production of different components of the pseudoscalar $P$ is
demonstrated via (a): SOZI process; (b) DOZI process; and (c)
glueball production. Similar processes apply to other $VP$ channels
as described in the text. }\protect \label{fig-2}
\end{center}
\end{figure}

For those isospin-conserved decays, i.e. $J/\psi(\psi^\prime) \to
\omega\eta,~\omega\eta^\prime,~\phi\eta,~\phi\eta^\prime,~\rho\pi$
and $K^*\bar K + c.c.$, the strong and EM decay processes are mixed.
Recalling that the antisymmetric tensor form is the only coupling
for $VVP$, we thus parameterize the strong decays in a way similar
to Refs.~\cite{Haber:1985cv,Seiden:1988rr,Zhao:2005im,Zhao:2007ze}.
Some basic quantities can be defined via Fig.~\ref{fig-2}: the
strength of non-strange singly OZI disconnected process
$g_{J/\psi}$; the parameter reflecting the SU(3) flavor symmetry
breaking effects $R$, and the parameter $r$ describing the relative
strength between the DOZI and SOZI transitions. The expressions for
the parameterized strong decay amplitudes are listed in
Table~\ref{tab-2}.

\begin{table}[ht]
\begin{tabular}{c|c}
\hline Decay channels  & Transition amplitude ${\cal M}=({\cal
M}_1+{\cal M}_2+{\cal M}_3)$
\\[1ex]
\hline  $\phi\eta$  & $g_{J/\psi(\psi^\prime)}R[\sqrt {2}r x_1
+R(1+r) y_1+ z_1]{\cal{F}}({\bf P})$
\\[1ex]
\hline  $\phi\eta^\prime$ & $g_{J/\psi(\psi^\prime)} R[\sqrt 2 r x_2
+
R (1+r) y_2 + z_2]{\cal{F}}({\bf P})$ \\[1ex]
\hline $\omega\eta$ &  $ g_{J/\psi(\psi^\prime)} [(1+2r)x_1+\sqrt
2Rr y_1 + \sqrt 2 z_1]{\cal{F}}({\bf P})$\\[1ex]
\hline $\omega\eta^\prime$  &  $ g_{J/\psi(\psi^\prime)}
[(1+2r)x_2+\sqrt
2R r y_2 + \sqrt 2 z_2]{\cal{F}}({\bf P})$\\[1ex]
\hline $\rho^0\pi^0$ &  $g_{J/\psi(\psi^\prime)} {\cal{F}}({\bf P})$ \\[1ex]
\hline $\rho^+\pi^-$ or $\rho^-\pi^+$  &  $g_{J/\psi(\psi^\prime)} {\cal{F}}({\bf P})$ \\[1ex]
\hline $ { K}^{* 0}\bar{K^0}$ or $\bar{K^{*0}} K^0$  & $g_{J/\psi(\psi^\prime)} R{\cal{F}}({\bf P})$ \\[1ex]
\hline $ K^{*+}K^-$  or $ K^{*-}K^+$ &  $g_{J/\psi(\psi^\prime)} R{\cal{F}}({\bf P})$  \\[1ex]
\hline
\end{tabular}
\caption{ General expressions for the transition amplitudes for
$J/\psi(\psi^\prime)\to VP$ via strong interactions. Parameter
$g_{J/\psi}$ and $g_{\psi^\prime}$ are proportional to the
charmonium wavefunctions at origin and have different values for
$J/\psi$ and $\psi^\prime$, respectively. For $\eta$ and
$\eta^\prime$, a glueball mixing is also considered in the
wavefunctions.} \label{tab-2}
\end{table}

In Ref.~\cite{Li:2007ky}, different treatments for glueball$-q{\bar
q}$ mixing are investigated, which are denoted by  Schemes I, II,
III. Since the glueball components in $\eta$ and $\eta^\prime$ are
rather small, the mixing effects will not change out results on the
strong decays of $J/\psi(\psi^\prime)\to VP$. Details about the
parameter definitions and mixings can be found in
Ref.~\cite{Li:2007ky}.

In order to take into account the size effects from the spatial
wavefunctions of the initial and final-state mesons, we apply the
commonly used form factor
\begin{eqnarray}
{\cal {F}}^2({\bf P}) \equiv |{\bf P}|^{2l}\exp ({-{\bf
P}^2/{8\beta^2}}),
\end{eqnarray}
where ${\bf P}$ and the $l$ are the three momentum and the relative
orbit angular momentum of the final-state mesons, respectively, in
the $J/\psi(\psi^\prime)$ rest frame. We adopt $\beta = 0.5 \ \mbox
{GeV}$, which is the same as
Refs.~\cite{Close:2005vf,Amsler:1995tu,Amsler:1995td,Close:2000yk}.

\begin{table}
 \begin{tabular}{|c|c|c|}
 \hline
& $g_{J/\psi}(\times 10^{-3})$ & $g_{\psi^\prime}(\times 10^{-3})$ \\
 \hline
Scheme-I &
$18.66\pm 0.63$ & $2.22 \pm 0.49$ \\
\hline
Scheme-II &
$18.45\pm 0.70$ & $2.54 \pm 0.44$ \\
\hline
Scheme-III & $17.52\pm 0.66$ & $2.57 \pm 0.42$ \\
\hline
 \end{tabular}
\caption{Extracted coupling strengths for the SOZI transitions for
three different $\eta-\eta^\prime-$glueball mixing schemes with a
constructive mode for the EM amplitudes~\protect\cite{Li:2007ky}.
}\label{tab-3}
 \end{table}

In Table~\ref{tab-3}, we list the values of the strong coupling
strengths $g_{J/\psi}$ and $g_{\psi^\prime}$ which are extracted by
overall fittings to the isospin-conserved decay channels of
$J/\psi(\psi^\prime)\to VP$ data including the EM transitions
determined in the previous section. The predominant feature is that
both values are stable and insensitive to the
$\eta-\eta^\prime-$glueball mixing schemes. In case of the absence
of the EM contributions, the ``$12\%$ rule" fraction should be
proportional to $(g_{J/\psi}/g_{\psi^\prime})^2$. By taking the
average of the squared values of Table~\ref{tab-3}, we obtain
$(g_{J/\psi}/g_{\psi^\prime})^2\simeq 1.8\%$ which is much less than
the expectation of the ``12\% rule", but larger than the
experimental data, $\sim (0.2\pm 0.1)\%$.

This is not at all a trivial outcome. Several points can be learned
here: i) The suppression to the $\psi^\prime$ strong decay coupling
is not exclusively on $\psi^\prime\to \rho\pi$. Such a suppression
is an overall effect on all the exclusive decays. ii) Due to the
suppression on the strong decay coupling of the $\psi^\prime$, the
EM transition amplitudes become compatible with the strong decay
amplitudes with which the interferences produce deviations from
naive expectations based on single transition mechanism. To be more
specific, due to the interference, the $\rho\pi$ decay is further
suppressed, i.e. causes the so-called ``$\rho\pi$ puzzle". The
neutral $K^{*0}\bar{K^0}+c.c.$ has larger branching ratio than the
charged one $K^{*+}K^-+c.c.$~\cite{pdg2006}. iii) As shown in
Fig.~\ref{fig-2}, the DOZI transitions contribute to the isoscalar
channels. This suggests that the exclusive decays have different
features compared with the inclusive one from which the ``12\% rule"
is embedded.

In Table~\ref{tab-4}, we list the branching ratio fractions for
those isospin-conserved channels and compare them with the
experimental data. Within the experimental uncertainties, our
results are in good agreement with the data. We also show the
branching ratio fractions for exclusive EM transitions, and again,
one can see that the ``12\% rule" is reasonably respected for
exclusive transitions.

 \begin{table}[ht]
 \begin{tabular}{cccccc}
 \hline
 Decay channels     & $R^{VP}(\%)$  & Scheme I(\%) & Scheme II(\%) & Scheme III(\%)   & Exp.data (\%)  \\[1ex]\hline
 $\rho^0\pi^0$      & 8               & 0.12         & 0.16         & 0.20           & *** \\[1ex]
 $\rho\pi$          & 8               & 0.12         & 0.15         & 0.19           & $0.2\pm 0.1$ \\[1ex]
 $\omega\eta$       & 8               & 0.40         & 0.35         & 0.28           & $< 0.6\pm 0.1$ \\[1ex]
 $\omega\eta^\prime$& 8               & 5.33         & 0.11         & 0.29           & $18.5\pm 13.2$ \\[1ex]
 $\phi\eta$         & 10              & 2.78         & 2.93         & 3.30           & $4.1\pm 1.6$ \\[1ex]
 $\phi\eta^\prime$  & 10              & 5.00         & 5.34         & 8.86           & $8.7\pm 5.5$ \\[1ex]
 $K^{*+}K^-+c.c.$   & 9               & 0.45         & 0.41         & 0.39           & $0.4\pm 0.2$ \\[1ex]
 $K^{*0}K^0+c.c.$   & 9               & 2.74         & 2.79         & 2.67           & $2.7\pm 0.7$
 \\[1ex]\hline
 \end{tabular}
\caption{  Branching ratio fractions for $\psi^\prime\to\gamma^*\to
VP$ over $J/\psi\to\gamma^*\to VP$ for different
$\eta$-$\eta^\prime-$ glueball with a MP-C form factor.  $R^{VP}$
denotes the ratios with exclusive EM transitions.  The last column
is extracted from the experimental date~\cite{pdg2006}. The stars
``***" in $\rho^0\pi^0$ channel denotes the unavailability of the
data. } \label{tab-4}
\end{table}


In brief, the parametrization identifies the mechanism which not
only causes puzzle in $\rho\pi$ channel, but also plays a role in
other $VP$ channels. since the EM transitions in $\rho\pi$ and
$K^*\bar{K}+c.c.$ are relatively large due to large couplings for
$\rho\pi\gamma$ and $K^*K\gamma$, interferences between the
suppressed strong decay amplitudes and enhanced EM amplitudes
produce significant deviations from the expectation of ``12\%
rule"~\cite{Zhao:2006gw,Li:2007ky}.

\section{Step 3: Mechanism suppressing the strong decay amplitudes
of $\psi^\prime\to VP$}

Now, the last bit of the whole scenario comes to the point, ``Why,
and how the strong decay coupling $g_{\psi^\prime}$ is suppressed?"
In order to demonstrate this, we express the decay amplitudes as
 \bea
{\cal M}_{J/\psi} &= & \frac{1}{M_{J/\psi}}(g_{J/\psi}
+g^{em}_{J/\psi}e^{i\delta_{J/\psi}}
) \epsilon_{\mu\nu\alpha\beta}\partial^\mu
V_{J/\psi}^\nu\partial^\alpha V_2^\beta P \nonumber\\
{\cal M}_{\psi^\prime} &= &
\frac{1}{M_{\psi^\prime}}(g_{\psi^\prime} + g^{em}_{\psi^\prime}
e^{i\delta_{\psi^\prime}}
) \epsilon_{\mu\nu\alpha\beta}\partial^\mu
V_{\psi^\prime}^\nu\partial^\alpha V_2^\beta P \ ,
 \eea
which again benefits from the property of the antisymmetric tensor
coupling among $VVP$ fields. In the above equation, $g_{J/\psi}$ and
$g_{\psi^\prime}$ are real numbers fixed by Step 2~\cite{Li:2007ky},
while $g^{em}_{J/\psi}$ and $g^{em}_{\psi^\prime}$ are the EM
couplings fixed by Step 1 with relative phase angles
$\delta_{J/\psi}$ and $\delta_{\psi^\prime}$ fixed in Step 2.
Detailed discussions on the phase angles can be found in
Ref.~\cite{Li:2007ky}, of which the values can be compared with
those from Ref.~\cite{Wang:2003hy}.

Since any possible mechanism must contribute to the coupling, we can
decompose the strong couplings as
 \bea\label{iml-correct}
g_{J/\psi} & \equiv & g_{J/\psi}^{pQCD} + g_{J/\psi}^{loop} \equiv
g_{J/\psi}^{pQCD}(1+q_{J/\psi})
\nonumber\\
g_{\psi^\prime} & \equiv & g_{\psi^\prime}^{pQCD} +
g_{\psi^\prime}^{loop} \equiv g_{\psi^\prime}^{pQCD}
(1+q_{\psi^\prime}) \ ,
 \eea
where $g_{J/\psi}^{pQCD}$ and $g_{\psi^\prime}^{pQCD}$ are couplings
given by pQCD power counting, while $g_{J/\psi}^{loop}$ and
$g_{\psi^\prime}^{loop}$ are given by an additional mechanism due to
intermediate meson loop transitions; quantities $q_{J/\psi}$ and
$q_{\psi^\prime}$ are the ratios of those two couplings for $J/\psi$
and $\psi^\prime$, respectively.  Qualitatively, suppression of the
$g_{\psi^\prime}$ coupling implies that there exist large
cancelations between $g_{\psi^\prime}^{pQCD}$ and
$g_{\psi^\prime}^{loop}$ while in $J/\psi$ decays effects from
$g_{J/\psi}^{loop}$ may not be significant.

Quantitative results supporting this require an explicit calculation
of both $g_{J/\psi}^{pQCD}$ and $g_{\psi^\prime}^{pQCD}$, for which
QCD models have been pursued in the literature. More or less, they
respect the ``12\% rule" since they probe the charmonium
wavefunctions at origin. The inclusion of intermediate meson loop
transitions will introduce corrections to the couplings via the
non-vanishing $q_{J/\psi}$ and $q_{\psi^\prime}$ in
Eq.~(\ref{iml-correct}). It is worth noting that the couplings from
the IML can be different for different decay channels. In
particular, for $\rho\pi$ channel, it turns that $|q_{J/\psi}| <
|q_{\psi^\prime}|$.

This relation again is not trivial at all. It further narrows down
the mechanism that causes the deviations from the pQCD power
counting, and also put a constraint on its behavior. As follows,
instead of providing detailed calculations for the loops, we
summarize the main features about the intermediate meson loop
transitions and detailed numerical results will be reported
later~\cite{li-zhao-forthcoming}:


I) Since both $J/\psi$ and $\psi^\prime$ are below the open charm
threshold, the intermediate meson loops will contribute to the real
part of the couplings. This feature not only justifies the
parametrization scheme in Step 2, but also makes the decomposition
of the strong couplings in Eq.~(\ref{iml-correct}) physically
meaningful.

II) Since the $\psi^\prime$ has a mass which is closer to the open
$D\bar{D}$ threshold, its amplitude via the $D\bar{D}$ loop will be
qualitatively larger than $J/\psi$ due to near-threshold effects.

III) Similar behavior due to intermediate $D\bar{D}(D^*)$ and
$D\bar{D^*}(D)$ loops also shows up in a coherent study of $J/\psi$
and $\psi^\prime\to \gamma\eta_c$ and $\psi^\prime\to
\gamma\eta_c^\prime$~\cite{Li:2007xr}.

IV) Light intermediate meson loops are strongly suppressed due to
large off-shell effects.

These features are consistent with  a recent study of the
``unquenched" effects arising from meson loops in
Ref.~\cite{barnes}, where it was shown that the intermediate meson
loops still play an important role within charmonium states below
the $D\bar{D}$ open threshold.

\section{Summary}

In this proceeding, we carry out a systematic analysis of the
problem of ``$\rho\pi$ puzzle", and clarify it on a more general
ground. We show that the EM transitions play an important role in
understanding the underlying mechanisms, which can be constrained by
the isospin-violating channels. It thus allows us to separate out
the EM amplitudes in those isospin-conserved channels.

The nature of the $VVP$ coupling as an antisymmetric tensor is also
a key for disentangling the problem since whatever the mechanisms
are for the transition, their contributions will simply be a
correction to the coupling form factor. This allows a
parametrization for the strong decay transitions in
$J/\psi(\psi^\prime)\to VP$. The result shows that there exists an
overall suppression on the strong decay amplitudes for
$\psi^\prime\to VP$. Because of such a suppression, the strong decay
amplitudes in some of those channels, such as $\rho\pi$ and
$K^*\bar{K}+c.c.$, become compatible with the EM transition
amplitudes, with which the interferences produce significant
deviations from the expectation of pQCD power counting rule. We then
identify that the suppression on the strong decay amplitudes is
originated from intermediate meson loop transitions, such as
$D\bar{D}(D^*)$, etc. In particular, the $\psi^\prime$ is closer to
the open $D\bar{D}$ threshold than $J/\psi$. As a result, it will
experience much larger threshold effects, and in this case a
destructive interference.

Such effects should be more general, hence may show up in other
decay channels. We point out that a larger intermediate meson loop
contribution originated from the same mechanism is also found in the
study of $J/\psi(\psi^\prime)\to\gamma\eta_c$ and
$\psi^\prime\to\gamma\eta_c^\prime$ as an important mechanism
interfering with the NRQCD leading amplitudes~\cite{Li:2007xr}.
Nevertheless, the study of ``unquenched" effects in the charmonium
spectrum also gives rise to the importance of the intermediate meson
loops~\cite{barnes}.

In brief, we clarify that the ``$\rho\pi$ puzzle" is not a single
problem with the $\rho\pi$ channel. Instead, it is a rather general
observation for all decay channels of $J/\psi(\psi^\prime)\to VP$.
In exclusive decays, multi-mechanisms can easily break down the pQCD
power counting due to interference. In this sense, the so-called
``$\rho\pi$ puzzle" should not be surprising. It turns to be more
interesting to us that the systematics arising from a coherent study
of all those $VP$ decay channels can provide us with much deeper
insights into the underlying dynamics. We expect that more precise
measurement of those isospin-violating decay branching ratios at
BESIII will help solve this long-standing
problem~\cite{Asner:2008nq,Li:2008ey}.

\section*{Acknowledgement}

The authors thank K.T. Chao, X.Q. Li, C.Z. Yuan, and B.S. Zou for
useful discussions on some of the relevant issues. This work is
supported, in part, by the National Natural Science Foundation of
China (Grant No.10675131 and 10491306), Chinese Academy of Sciences
(KJCX3-SYW-N2), and the U.K. EPSRC (Grant No. GR/S99433/01). C.H.C
is supported by National Natural Science Foundation of China (Grant
No.10547001 and No.90403031).

\end{document}